\documentclass[12pt]{article}
\usepackage{graphicx}
\newcommand{\posthole}[2]{\vspace{5cm}}

\font\tenbf=cmbx10
\font\tenrm=cmr10
\font\tenit=cmti10
\font\elevenbf=cmbx10 scaled\magstep 1
\font\elevenrm=cmr10 scaled\magstep 1
\font\elevenit=cmti10 scaled\magstep 1

\renewenvironment{thebibliography}[1]
 { \elevenrm 
   \begin{list}{\arabic{enumi}.}
    {\usecounter{enumi} \setlength{\parsep}{0pt}
     \setlength{\itemsep}{3pt} \settowidth{\labelwidth}{#1.}
     \sloppy
    }}{\end{list}}

\newcommand{\beqa}{\begin{eqnarray}}
\newcommand{\eeqa}{\end{eqnarray}}
\def\lag{Lagrangian}
\def\ra{\rightarrow}

\newcommand{\RMP}[3]{{\elevenit Rev. Mod. Phys.} {\elevenbf  #1}, #2 (19#3)}
\newcommand{\PR}[3]{{\elevenit Phys. Rev.} {\elevenbf #1}, #2 (19#3)}
\newcommand{\PL}[3]{{\elevenit Phys. Lett.} {\elevenbf #1}, #2 (19#3)}

\newcommand{\PRL}[3]{{\elevenit Phys. Rev. Lett.} {\elevenbf #1}, #2 (19#3)}
\newcommand{\NP}[3]{{\elevenit Nucl. Phys.} {\elevenbf #1}, #2 (19#3)}
\newcommand{\ZP}[3]{{\elevenit Zeit. Phys.} {\elevenbf #1}, #2 (19#3)}
\newcommand{\con}[3]{{\elevenbf #1}, #2 (19#3)}

\newcommand{\skipblk}[1]{}                                                      
\def\bqa{\begin{eqnarray}}                                                      
\def\eqa{\end{eqnarray}}

\newcommand{\etal}{{\elevenit et al., }}

\newcommand{\ie}{{\elevenit i.e., }}

\newcommand{\x}{\mbox{$\times$}}

\newcommand{\beq}{\begin{equation}}                                             
\newcommand{\eeq}{\end{equation}}

\newcommand{\p}[1]{\mbox{$10^{#1}$}}

\def\mxth{\mathsurround=0pt }
\def\xversim#1#2{\lower2.pt\vbox{\baselineskip0pt \lineskip-.5pt
  \ialign{$\mxth#1\hfil##\hfil$\crcr#2\crcr\sim\crcr}}}

\def\beqa{\begin{eqnarray}}
\def\eeqa{\end{eqnarray}}

\def\etal{{\it et al.,\ } }

\def\mxth{\mathsurround=0pt }
\def\xversim#1#2{\lower2.pt\vbox{\baselineskip0pt \lineskip-.5pt
  \ialign{$\mxth#1\hfil##\hfil$\crcr#2\crcr\sim\crcr}}}
\def\grsim{\mathrel{\mathpalette\xversim >}}
\def\CP{$CP$}

\def\lsim{\mathrel{\mathpalette\xversim <}}

\def\x{\times}
\def\ra{\rightarrow}

\def\L{{\cal L}}

\def\lag{Lagrangian}
\def\her{Hermitian}

\def\beq{\begin{equation}}
\def\eeq{\end{equation}}

\begin{document}
% \ifpdf
%\DeclareGraphicsExtensions{.jpg,.pdf,.mps,.png}
%\else
%\DeclareGraphicsExtensions{.eps,.ps}
% \fi

\begin{center}
%{\tenbf STRUCTURE OF THE STANDARD MODEL\\}
{\tenbf STRUCTURE OF THE STANDARD MODEL\footnote{Reprinted from
   {\it Precision Tests of the Standard Electroweak Model}, ed.
   P. Langacker (World, Singapore, 1995).}\\}
\vglue 1.0cm
{\tenrm PAUL LANGACKER \\}
\baselineskip=13pt
{\tenit Department of Physics, University of Pennsylvania,\\
 Philadelphia, Pennsylvania, USA 19104-6396 \\}
\end{center}

\vglue 0.6cm
\tableofcontents
\baselineskip=14pt
\elevenrm

\vglue 0.6cm
\section{\elevenbf The Standard Model Lagrangian}
\vglue 0.4cm

The standard model~\cite{ss1} is a gauge theory~\cite{ss1a} of the microscopic
interactions. The strong interaction part (QCD~\cite{QCD}) is described by the
Lagrangian \beq \L_{SU_3} = - \frac{1}{4} F^i_{\mu\nu} F^{i\mu\nu} + \sum_r
\bar{q}_{r\alpha} i \not{\!\!D}^\alpha_\beta \,q^\beta_r, \label{eq1:c20}
\eeq
where $g_s$ is the QCD gauge coupling
constant,
\beq F^i_{\mu\nu} = \partial_\mu G^i_\nu - \partial_\nu G^i_\mu -
g_s f_{ijk}\; G_\mu^j\; G_\nu^k  \label{eq1:c21} \eeq
is the field strength tensor for the gluon
fields $G^i_\mu, \; i = 1, \cdots, 8$,  and the structure constants $f_{ijk}$
 $ (i, j, k = 1, \cdots, 8)$ are
defined by
\beq [\lambda^i, \lambda^j] = 2 i f_{ijk} \lambda^k, \eeq
where the $SU_3$ $\lambda$ matrices are defined in
Table~\ref{table1:000b}.
    The $F^2$ term leads to three and four-point gluon
self-interactions. The % in Figure~\ref{figch4}.  The
second term in $\L_{SU_3}$ is the gauge covariant derivative for the
quarks: $q_r$ is the $r^{\rm th}$ quark flavor, $\alpha, \beta = 1,2,3$ are
color indices, and \beq D^\alpha_{\mu \beta} = (D_\mu)_{\alpha \beta} =
\partial_\mu \delta_{\alpha \beta} + i g_s \;G^i_\mu\; L^i_{\alpha\beta},
\label{eq1:c22} \eeq
where the quarks transform according to the triplet representation matrices
$L^i ={\lambda^i}/{2}$.
The color interactions are diagonal in the flavor
indices, but in general change the quark colors. They are purely vector
(parity conserving). There are no bare mass terms for the quarks in
(\ref{eq1:c20}). These would be allowed by QCD alone, 
but are forbidden by the chiral
symmetry of the electroweak part of the theory.  The quark masses will be
generated later by spontaneous symmetry breaking.  There are in addition
effective ghost and gauge-fixing terms
which enter into the quantization of both the $SU_3$ and
electroweak \lag s, and there is the possibility of adding an (unwanted) term
which violates $CP$ invariance.

\begin{table}\centering
\small \def\baselinestretch{1} \normalsize
\begin{tabular}{|rrr|} \hline
$\lambda^i = \left( \begin{array}{cc} \tau^i & 0 \\ 0 & 0
\end{array} \right),$ & $i = 1,2,3$ &\\
$\lambda^4 = \left( \begin{array}{ccc} 0 & 0 & 1 \\ 0 & 0 & 0 \\
1 & 0 & 0 \end{array} \right)$ & &
\ \ \ $\lambda^5 = \left( \begin{array}{lcr} 0 & 0 & -i \\ 0 & 0
& 0 \\ i & 0 & 0 \end{array} \right)$ \\
$\lambda^6 = \left( \begin{array}{ccc} 0 & 0 & 0 \\ 0 & 0 & 1 \\
0 & 1 & 0 \end{array} \right)$ & &
$\lambda^7 = \left( \begin{array}{lcr} 0 & 0 & 0 \\ 0 & 0 & -i
\\ 0 & i & 0 \end{array} \right)$ \\
$\lambda^8 = \frac{1}{\sqrt{3}} \left( \begin{array}{lcr} 1 & 0 &
0 \\ 0 & 1 & 0 \\ 0 & 0 & -2 \end{array} \right)$ & & \\ \hline
\end{tabular}
\caption{The $SU_3$ matrices.}
\label{table1:000b}
\end{table}

%\begin{figure}
%\small \def\baselinestretch{1} \normalsize
%\vspace{12cm}
%\caption{Interactions in QCD.}
%\label{figch4}
%\end{figure}
 
The electroweak theory is based on the $SU_2 \x U_1$ \lag\ ~\cite{ew}
\beq \L_{SU_2 \x U_1} = \L_{\rm gauge} + \L_\varphi + \L_f + \L_{\rm
Yukawa}. \label{eqch10b} \eeq
The gauge part is
\beq \L_{\rm gauge} = - \frac{1}{4} F^i_{\mu \nu} F^{\mu \nu i} -
\frac{1}{4} B_{\mu \nu} B^{\mu \nu}, \label{eqch11b} \eeq
where $W^i_\mu, \; i = 1, \; 2, \; 3$ and
$B_\mu$  are respectively the $SU_2$ and $U_1$ gauge fields,
with field  strength tensors 
\begin{eqnarray} B_{\mu \nu} &=& \partial_\mu B_\nu - \partial_\nu
B_\mu \nonumber \\
F_{\mu \nu} &=& \partial_\mu W_\nu^i - \partial_\nu W_\mu^i - g
\epsilon_{ijk} W^j_\mu W^k_\nu, \label{eqch12a} \end{eqnarray}
where $g (g')$ is the $SU_2$ $(U_1)$ gauge coupling and $\epsilon_{ijk}$ is
the totally antisymmetric symbol.
 The $SU_2$ fields
have three and four-point self-interactions.
%analogous to those in Figure~\ref{figch4}.
$B$ is a $U_1$ field associated with the
weak hypercharge 
$ Y = Q - T_3$, where $Q$ and $T_3$ are respectively the electric charge
operator and the third component of weak $SU_2$. It has no
self-interactions. The $B$ and $W_3$ fields will eventually mix to form the
photon and $Z$~boson.
 
The scalar part of the \lag\ is
\beq \L_\varphi = (D^\mu \varphi)^{\dag} D_\mu \varphi - V(\varphi),
\label{eqch13b} \eeq
where $\varphi = \left(\begin{array}{c} \varphi^+ \\ \varphi^0 \end{array}
\right)$ is a complex Higgs scalar, which is a doublet under $SU_2$ with $U_1$
charge $Y_\varphi = + \frac{1}{2} $.  The gauge covariant derivative is
\beq D_\mu \varphi = \left( \partial_\mu + i g \frac{\tau^i}{2}
W_\mu^i + \frac{i g'}{2} B_\mu \right) \varphi, \label{eqch14b} \eeq
where the $\tau^i$ are the Pauli matrices.
   The square of the covariant derivative leads to three and four-point
interactions between the gauge and scalar fields~\cite{ss1}.
 
$V(\varphi)$ is the Higgs potential.  The combination of $SU_2 \x U_1$
invariance and renormalizability restricts $V$ to the form
\beq V(\varphi) = + \mu^2 \varphi^{\dag}\varphi + \lambda (\varphi^{\dag}
\varphi)^2. \label{eq15} \eeq
For $\mu^2 < 0$ there will be spontaneous symmetry breaking.  The
$\lambda$ term describes a quartic  self-interaction
between the scalar fields. Vacuum stability requires $\lambda > 0$.
 
The fermion term is
\beq \L_F = \sum^F_{m = 1} \left( \bar{q}^0_{mL} i
\not\!\!D q^0_{mL} + \bar{l}^0_{mL} i \not\!\!D l^0_{mL}
 + \bar{u}^0_{mR} i \not\!\!D u^0_{mR} +
\bar{d}^0_{mR} i \not\!\!D d^0_{mR} + \bar{e}^0_{mR} i \not\!\!D
e^0_{mR} \right). \label{eqch16} \eeq
In (\ref{eqch16}) $m$ is the family index, $F \ge 3$ is the number
of families, and $L(R)$ refer to the left (right) chiral projections
$\psi_{L(R)} \equiv (1 \mp \gamma_5) \psi/2$. The left-handed quarks
and leptons 
\beq
q^0_{mL}= \left( \begin{array}{c} u^0_m \\ d^0_m \end{array}
\right)_L \ \ \ \ \ l^0_{mL} = \left( \begin{array}{c} \nu^0_m \\ e^{-0}_m
\end{array} \right)_L
\eeq
transform as $SU_2$ doublets, while the right-handed fields
$u^0_{mR}, \; d^0_{mR}$, and  $e^{-0}_{mR}$ are singlets. 
Their $U_1$ charges are  $Y_{q_L} = \frac{1}{6}, \; Y_{l_L} =-\frac{1}{2}, \;
Y_{\psi_R} = q_\psi$. The superscript $0$ refers to the weak 
eigenstates, \ie fields
transforming according to definite $SU_2$ representations.  They may
be mixtures of mass eigenstates (flavors).
The quark color indices $\alpha = r,
\; g, \; b$ have been suppressed. 
The gauge covariant derivatives are
%\begin{eqnarray}
%D_\mu q^0_{mL} &=& \left( \partial_\mu + \frac{i g}{2} \tau^i W^i_\mu
%+ i\frac{ g'}{6} B_\mu \right) q^0_{mL} \nonumber \\
%D_\mu l^0_{mL} &=& \left( \partial_\mu + \frac{i g}{2} \tau^i W^i_\mu
%- i\frac{ g'}{2} B_\mu \right) l^0_{mL} \nonumber \\
%D_\mu u^0_{mR} &=& \left( \partial_\mu + i\frac{2}{3}g' B_\mu \right)
%u^0_{mR} \nonumber \\
%D_\mu d^0_{mR} &=& \left(\partial_\mu -i\frac{g'}{3}B_\mu  \right)
%d^0_{mR} \nonumber \\
%D_\mu e^0_{mR} &=& \left( \partial_\mu - i g' B_\mu     \right)
%e^0_{mR}\;\; , \label{eqch17} \end{eqnarray}
\begin{equation}  \begin{array}{lclclcl}
D_\mu q^0_{mL} &=& \left( \partial_\mu + \frac{i g}{2} \tau^i W^i_\mu
+ i\frac{ g'}{6} B_\mu \right) q^0_{mL}
&  \ \ \ \ \ &
D_\mu u^0_{mR} &=& \left( \partial_\mu + i\frac{2}{3}g' B_\mu \right)
u^0_{mR}  \\
D_\mu l^0_{mL} &=& \left( \partial_\mu + \frac{i g}{2} \tau^i W^i_\mu
- i\frac{ g'}{2} B_\mu \right) l^0_{mL}
&  \ \ \ \ \ &
D_\mu d^0_{mR} &=& \left(\partial_\mu -i\frac{g'}{3}B_\mu  \right)
d^0_{mR}  \\
& &  &  \ \ \ \ \ &
D_\mu e^0_{mR} &=& \left( \partial_\mu - i g' B_\mu     \right)
e^0_{mR},
\end{array}  \label{eqch17} \end{equation}
from which one can read off the gauge interactions between the $W$ and $B$
and the fermion fields.
 The different transformations of the $L$ and $R$ fields (\ie  the symmetry
is chiral) is the origin of parity violation in the electroweak sector. 
The chiral symmetry also forbids
any bare mass terms for the fermions.
 
The last term in (\ref{eqch10b}) is 
\beq -L_{\rm Yukawa} = \sum^F_{m,n =1} \left[
\Gamma^u_{mn} \bar{q}^0_{mL} \tilde{\varphi} u^0_{mR} + \Gamma^d_{mn}
\bar{q}^0_{mL} \varphi d^0_{nR} 
+ \Gamma^e_{mn} \bar{l}^0_{mn} \varphi e^0_{nR} \right] +
{\rm H.C.}, \label{eqch18} \eeq
where the matrices $\Gamma_{mn}$  describe the  Yukawa
couplings between the single Higgs doublet, $\varphi$, and the various
flavors $m$ and $n$ of quarks and leptons.  One needs representations of
Higgs fields with
 $Y = \frac{1}{2}$ and $-\frac{1}{2}$ to
give masses to the down quarks, the electrons, and the up
quarks.    The representation $\varphi^{\dag}$ has $Y = - \frac{1}{2}$,
but transforms as the $2^*$ rather than the 2.  However, in
                   $SU_2 $              the $2^*$
representation is related to the 2 by a similarity
transformation, and
$\tilde{\varphi} \equiv i \tau^2 \varphi^{\dag} = \left( \begin{array}{c}
\varphi^{0^{\dag}} \\ - \varphi^- \end{array} \right)$ transforms as a 2
with $Y_{\tilde{\varphi}}=-\frac{1}{2}$. All
of the masses can therefore be generated with a single Higgs doublet if
one makes use of both $\varphi$ and $\tilde{\varphi}$.  The fact that the
fundamental and its conjugate are equivalent does not generalize to higher
unitary groups.  Furthermore, in supersymmetric extensions of the standard
model  the supersymmetry forbids the use of a single Higgs doublet in both
ways in the \lag, and one must add a second Higgs
doublet. Similar statements apply to most theories with an
additional $U_1$ gauge factor, \ie a heavy $Z'$ boson.

\vglue 0.6cm
\section{\elevenbf Spontaneous Symmetry Breaking}
\vglue 0.4cm
 
Gauge invariance (and therefore renormalizability) does not allow
mass terms in the \lag \ for the gauge bosons or for chiral fermions.
Massless gauge bosons are not acceptable for the weak interactions, 
which are known to be short-ranged. Hence, the gauge invariance must be
broken spontaneously~\cite{ssb}, which preserves
the renormalizability~\cite{J001}. The idea is simply that the lowest energy
(vacuum) state does not respect the gauge symmetry and induces
effective masses for particles propagating through it.
 
Let us introduce the
complex vector
\beq v = \langle 0 | \varphi | 0 \rangle = {\rm
constant}, \label{eqch20} \eeq
which has components that are  the vacuum
expectation values of the various complex scalar fields.
$v$ is determined  by rewriting the Higgs potential
      as a function of $v$, $V(\varphi) \ra V(v)$, and choosing $v$
such that $V$ is minimized.  That is, we
interpret $v$   as the lowest energy
solution  of the classical  equation of motion\footnote{It suffices to
consider constant $v$ because
     any space or time dependence $\partial_\mu v$
would increase the energy of the solution. Also, one can take
$\langle 0 | \psi | 0 \rangle = \langle 0 | A_\mu | 0
\rangle = 0, \label{eqch21}$
because any non-zero vacuum values would
violate Lorentz invariance. These extensions are involved in (higher
energy) topological defects, such as monopoles, strings, domain
walls, and textures.}.         
The quantum theory is
 obtained by
considering fluctuations around this classical minimum,  $\varphi = v +
\varphi'$.
 
\begin{figure}
\small \def\baselinestretch{1} \normalsize
%\vspace{6cm}
\centering
\includegraphics*[scale=1.0]{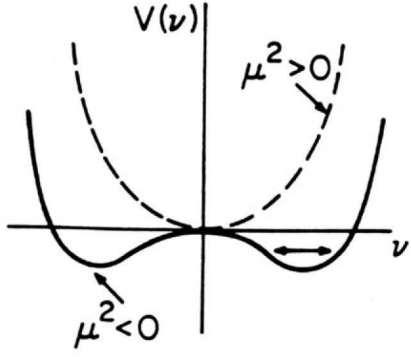}
\caption{The Higgs potential $V(\nu)$ for
$\mu^2 > 0$ (dashed line) and $\mu^2 < 0$ (solid line).}
\label{figure11}
\end{figure}

The single complex Higgs
doublet in  the standard model can be rewritten in a Hermitian basis as
\beq \varphi = \left( \begin{array}{c} \varphi^+ \\ \varphi^0
\end{array} \right) = \left( \begin{array}{c} \frac{1}{\sqrt{2}} (
\varphi_1 - i \varphi_2)        \\       \frac{1}{\sqrt{2}}(\varphi_3- i
\varphi_4  \end{array} \right), \label{eqch22} \eeq
where $\varphi_i = \varphi_i^{\dag}$
represent four \her\ fields.  In this new basis the Higgs potential
becomes
\beq V(\varphi) = \frac{1}{2} \mu^2 \left( \sum^4_{i=1} \varphi^2_i \right) +
\frac{1}{4} \lambda \left( \sum^4_{i=1} \varphi^2_i \right)^2
, \label{eqch23} \eeq
which is clearly $O_4$ invariant.  Without loss
of generality we can choose the axis   in this four-dimensional space
so that $\langle 0| \varphi_i |0 \rangle = 0, \;\ i = 1, 2, 4 $
and $\langle 0 | \varphi_3 | 0 \rangle = \nu$. Thus,
\beq V (\varphi) \ra V(v) = \frac{1}{2} \mu^2 \nu^2 + \frac{1}{4}
\lambda \nu^4, \label{eqch24} \eeq
which must be minimized with respect to $\nu$.  Two important cases
are illustrated in Figure~\ref{figure11}.  For $\mu^2 > 0$ the minimum
occurs at $\nu = 0$.  That is, the vacuum is empty space and $SU_2 \x
U_1$ is unbroken at the minimum.  On the other hand, for $\mu^2< 0$
the $\nu = 0$ symmetric point is unstable, and the minimum occurs at
some nonzero value of $\nu$ which breaks the $SU_2 \x U_1$ symmetry.
The point is found by requiring
\beq V' (\nu) = \nu (\mu^2 + \lambda \nu^2) = 0,\label{eqch25} \eeq
which has the solution
$\nu = \left(      {-\mu^2}/{\lambda} \right)^{1/2} $
at the minimum.  (The solution for $-\nu$ can also be transformed into
this standard form by an appropriate $O_4$ transformation.)
The dividing point $\mu^2 = 0$ cannot be treated classically.  It
is necessary to consider the one loop corrections to the potential, in
which case it is found that the symmetry is again spontaneously
broken~\cite{9h}.
 
We are interested in the case $\mu^2 < 0$, for which the Higgs doublet
is replaced, in first approximation, by its classical value
$ \varphi \ra \frac{1}{\sqrt{2}} \left( \begin{array}{c} 0 \\ \nu
\end{array} \right) \equiv v$.
The generators $L^1$, $L^2$, and $L^3 - Y$ are spontaneously
broken ({\em e.g.,} $L^1 v \neq 0$).  On the other hand, the vacuum carries no
electric charge ($Qv = (L^3 + Y) v = 0$),
so the $U_{1Q}$ of electromagnetism is not broken.  Thus,
the electroweak $SU_2 \x U_1$ group is spontaneously broken down,
$SU_2 \x U_{1Y} \ra U_{1Q}$.

To quantize around the classical vacuum, write $\varphi = v +
\varphi'$, where $\varphi'$ are quantum fields with zero vacuum
expectation value. To display
the physical particle content it is useful to rewrite the four \her\
components of $\varphi'$ in terms of a new set of variables using  the
Kibble transformation~\cite{9e}: \beq \varphi = \frac{1}{\sqrt{2}}
e^{i \sum \xi^i L^i} \left( \begin{array}{c} 0 \\ \nu + H \end{array}
\right). \label{eqch29} \eeq $H$ is a \her\ field which will turn out to be
the physical Higgs
scalar. If we
had been dealing with a spontaneously broken global symmetry the three \her\
fields $\xi^i$ would be the massless pseudoscalar Goldstone bosons~\cite{9i}
that are necessarily associated with broken symmetry generators.  However, in
a            gauge theory they disappear from the physical spectrum. To see
this      it is useful to go to the unitary gauge
\beq
\varphi \ra \varphi' = e^{-i \sum \xi^i L^i} \varphi =
\frac{1}{\sqrt{2}} \left( \begin{array}{c} 0 \\ \nu + H \end{array}
\right), \label{eqch30} \eeq
                           in which the Goldstone bosons
disappear.                          In this gauge, the scalar
covariant kinetic energy term takes the simple form
\begin{eqnarray} (D_\mu \varphi)^{\dag} D^\mu \varphi & = & \frac{1}{2} (0\;
\nu) \left[ \frac{g}{2} \tau^i W^i_\mu + \frac{g'}{2} B_\mu \right]^2
\left( \begin{array}{c} 0 \\ \nu \end{array} \right) 
+ H \; {\rm  terms} \nonumber \\
& \ra & M^2_W W^{+\mu} W^-_\mu +
\frac{M_Z^2}{2} Z^\mu Z_\mu + H \; {\rm  terms}, \label{eqch31}
\end{eqnarray}
where the kinetic energy and gauge interaction terms of the physical $H$
particle have been omitted. Thus,
 spontaneous symmetry breaking generates mass terms for the
$W$ and $Z$ gauge bosons
\begin{eqnarray} W^{\pm}&=&\frac{1}{\sqrt{2}} (W^1 \mp i W^2)\nonumber\\
Z &=& -\sin \theta_W B + \cos \theta_W W^3.\label{eqch33} \end{eqnarray}
The photon field
\beq A = \cos \theta_W B + \sin \theta_W W^3 \label{eqch34}
\eeq
remains massless.  The masses are
\beq M_W = \frac{g \nu}{2} \label{eqch34b} \eeq
and
\beq M_Z = \sqrt{g^2 + g^{\prime 2} } \frac{\nu}{2} = \frac{M_W}{\cos
\theta_W}, \label{eqch35} \eeq
where the weak angle is defined by $\tan\theta_W\equiv {g'}/{g} $.
One can think of the generation of masses as due to the fact that the
$W$ and $Z$ interact constantly with the condensate of scalar fields and
therefore acquire masses, in analogy with a photon propagating through a
plasma. The Goldstone boson has
disappeared from the theory  but has reemerged as the longitudinal degree
of freedom of a massive vector particle.

It will be seen below that  $G_F/\sqrt{2} \sim g^2/8 M^2_W$, where $G_F
= 1.16639(2) \x \p{-5}\ GeV^{-2}$ is the Fermi constant determined by the
muon lifetime. The weak scale $\nu$ is therefore 
\beq \nu = 2M_W/g \simeq
(\sqrt{2} G_F)^{-1/2} \simeq 246  \;GeV. \label{3-45a} \eeq
Similarly, $g = e/\sin \theta_W$, where $e$ is the electric charge of the
positron. Hence, to lowest order 
\beq
M_W = M_Z \cos \theta_W \sim \frac{(\pi \alpha/\sqrt{2} G_F)^{1/2}}{
\sin \theta_W}, \label{wztree} \eeq
where $\alpha \sim 1/137.036$ is the fine structure constant. Using $\sin^2
\theta_W \sim 0.23$ from neutral current scattering, one expects $M_W \sim
78\ GeV$, and $M_Z \sim 89\ GeV$. (These predictions are increased by
$\sim (2-3)\ GeV$ by loop corrections.)  
          The $W$ and $Z$ were discovered at CERN by the UA1~\cite{lac29}
           and UA2~\cite{lac30}
groups in 1983.  Subsequent measurements of their masses and other
properties have been in perfect agreement with the standard model
expectations (including the higher-order corrections), as is
described in the articles of by Schaile and Einsweiler.

After  symmetry breaking   the Higgs potential becomes
\beq V(\varphi) = - \frac{\mu^4}{4\lambda} - \mu^2 H^2 + \lambda \nu
H^3 + \frac{\lambda}{4} H^4.\label{eqch37} \eeq
The third and fourth terms  represent the cubic and
quartic interactions of the Higgs scalar.
The second term represents a (tree-level) mass 
\beq M_H = \sqrt{-2\mu^2} = \sqrt{2 \lambda} \nu . \label{eqch38} \eeq
The weak scale is given in (\ref{3-45a}), but the quartic Higgs coupling
$\lambda$ is unknown, so $M_H$ is not predicted. A priori, $\lambda$
could be anywhere
in the range $ 0 \leq \lambda < \infty $. 
 There is now an experimental lower limit $M_H \grsim
60$~GeV from LEP~\cite{9j}.  Otherwise, the decay $Z \ra Z^* H$ would have
been observed. (There are also theoretical lower limits on $M_H$ in the 0
-- 10~GeV range, depending on $m_t$, when higher-order corrections are
included~\protect\cite{9f}.) 
 
  There are also plausible theoretical
upper limits. If $\lambda > O (1)$
the theory becomes strongly coupled.
$(M_H > O (1$~TeV)).  There is not really anything wrong with strong
coupling a priori.  However, there are fairly convincing triviality
limits, which basically say that the running
quartic coupling would become infinite within the domain of validity
of the theory if $\lambda$ and therefore $M_H$ is too large.  If one
requires the theory to make sense to infinite energy, one may run
into problems\footnote{This is true for a  pure $\lambda H^4$ theory.
The presence of other interactions may eliminate the problems for
small $\lambda$.}
 with the increasing quartic coupling for any $\lambda$.
However, one only needs for  the theory to hold up to the next
mass scale $\Lambda$, at which point the standard model breaks down. In that
case~\cite{9f}, \beq
M_H < \left\{ \begin{array}{l} O (200) \;GeV, \; \Lambda \sim M_P\\
                               O (600) \;GeV, \; \Lambda \sim 2M_H
\end{array} \right. \label{eqch39} \eeq
The more stringent limit of $O(200)$~GeV obtains for $\Lambda$ of order of
the Planck scale $M_P = G_N^{-1/2} \sim 10^{19}$~GeV.  If one makes
the less restrictive assumption that the scale $\Lambda$ of new
physics can be small, one obtains a weaker limit.  Nevertheless, for
the concept of an elementary Higgs field to make sense one should
require that the theory be valid up to something of order of $2M_{H}$,
which implies that $M_H < O$(600)~GeV.  These limits may be relaxed
if there are other heavy particles in the theory.
 
The first term in (\ref{eqch37}) is the vacuum expectation value
\beq \langle 0| V|0 \rangle = - \mu^4 / 4 \lambda \label{eqch40} \eeq
of the Higgs potential when evaluated at the minimum.  This is a $c$-number
which has no significance for the microscopic interactions.
However, it assumes great importance when the theory is coupled to
gravity, because a constant energy density plays the role of a
cosmological constant~\cite{coscon}.  The cosmological
constant becomes
\beq \Lambda_{\rm cosm} = \Lambda_{\rm bare} + \Lambda_{\rm SSB} ,
\label{eqch41} \eeq
where $\Lambda_{\rm bare}$ is the primordial cosmological constant,
which can be thought of as the value of the energy of the vacuum in
the absence of spontaneous symmetry breaking. (Eqn.  (\ref{eq15})
implicitly assumed $\Lambda_{\rm bare} = 0$.) $\Lambda_{\rm SSB}$ is
the part generated by the Higgs mechanism:
\beq | \Lambda_{\rm SSB}| = 8 \pi G_N |\langle 0 | V| 0 \rangle |
\sim 10^{50}| \Lambda_{\rm obs} |.\label{eqch42} \eeq
It is some $10^{50}$ times larger than the observational upper limit
$\Lambda_{\rm obs}$.  This is clearly unacceptable. Technically, one
can solve the problem by adding a constant $+\mu^4/4 \lambda$ to $V$, so
that $V$ is equal to zero at the minimum ({\it i.e.,} $\Lambda_{\rm
bare} = 2 \pi G_N \mu^4/\lambda)$.  However, with our current
understanding there is no reason for $\Lambda_{\rm bare}$ and
$\Lambda_{\rm SSB}$ to be related; to have to invoke such an
incredibly fine-tuned cancellation to 50 decimal places is a major
unsatisfactory feature of the standard model.
 
The Yukawa interaction  in the unitary gauge  becomes
\begin{eqnarray} -L_{\rm Yukawa} &\ra& \sum^F_{m,n = 1} \bar{u}^0_{mL}
\Gamma^u_{mn}\left( \frac{\nu + H}{\sqrt{2}} \right) u^0_{mR} + (d,
e) {\rm \; terms \;\; + \;\; H.C.} \nonumber \\
&=& \bar{u}^0_L \left( M^u + h^u H \right) u^0_R + (d, e) \; {\rm
terms \;\; + H.C.,} \label{eqch43} \end{eqnarray}
where in the second form
$u^0_L = \left( u^0_{1L} u^0_{2L} \cdots u^0_{FL} \right)^T$
is an $F$-component column vector, with a similar definition for
$u_R^0$.   $M^u$ is an $F\x F$
fermion mass matrix
$M^u_{mn} = \Gamma^u_{mn} {\nu}/{\sqrt{2}}$
induced by spontaneous symmetry
breaking, and  $ h^u = M^u/\nu = {g M^u}/{2 M_W}$ is the Yukawa
coupling matrix.
 
In general $M$ is not diagonal, \her, or symmetric. To
identify the physical  particle content it is necessary to
diagonalize $M$ by separate unitary transformations $A_L$ and $A_R$ on the
left- and right-handed fermion fields.  (In the special case that $M^u$ is
\her\ one can take $A_L = A_R$).  Then, \beq A_L^{u \dag} M^u A^u_R = M^u_D
= \left( \begin{array}{ccc} m_u & 0 & 0 \\ 0 & m_c & 0 \\ 0 & 0 & m_t
\end{array} \right) \label{eqch44} \eeq
is a diagonal matrix with eigenvalues equal to the physical masses of
the charge $\frac{2}{3}$ quarks. Similarly, one
diagonalizes the down quark and charged lepton mass matrices by
\begin{eqnarray}
A^{d\dag}_L M^d A^d_R &=& M^d_D \nonumber \\
A^{e\dag}_L M^e A^e_R &=& M^e_D. \label{eqch45} \end{eqnarray}
In terms of these unitary matrices we can define mass eigenstate
fields
$u_L = A_L^{u\dag} u_L^0 = (u_L\ c_L\ t_L)^T$,  with analogous definitions
for $u_R = A_R^{u\dag} u_R^0$,
$d_{L,R} = A_{L,R}^{d\dag} d_{L,R}^0$, and $
e_{L,R} = A_{L,R}^{e\dag} e_{L,R}^0$.
Assuming the neutrinos are massless, their mass eigenstates are
arbitrary. It is convenient to define them in terms of the
charged lepton unitary transformation, $\nu_L = A_L^{e\dag} \nu_L^0$.
That is, we define
$\nu_e,\; \nu_\mu,\; \nu_\tau$  as the weak interaction partners of
the $e, \; \mu$, and $\tau$. Typical estimates of the  quark masses
are~\cite{9L} $
m_u=5.6 \pm 1.1 \;MeV,\  m_d=9.9 \pm 1.1 \;MeV,\ m_s=199 \pm 33 \;MeV,\ m_c=
1.35 \pm 0.05\; GeV,\ m_b \sim 4.7\; GeV,$ and
$ m_t > 131\; GeV$~\cite{d0} or $m_t = 174 \pm 16\; GeV$~\cite{CDF}. These are
the current masses: for QCD their effects are identical to bare masses in
the QCD \lag.  They should not be confused with the constituent masses of
order 300~MeV generated by the spontaneous breaking of chiral symmetry in
the strong interactions.  Including QCD renormalizations, the $u, \; d, \;
s$ and $c$ masses are running masses evaluated at
1~GeV$^2$, while $m_b$ and $m_t$ are pole masses.

Thus,
\beq L_{\rm Yukawa} = \sum_i \bar{\psi}_i \left( - m_i
-\frac{gm_i}{2M_W} H \right) \psi_i. \label{eqch48} \eeq   The
 coupling of the physical Higgs boson to the $i^{\rm th}$ fermion
is $gm_i/2M_W$, which is very small except for the top quark.  The coupling
is flavor-diagonal in the minimal model: there is just one Yukawa
matrix for each type of fermion, so the mass and Yukawa matrices are 
diagonalized by
the same transformations.
 In generalizations in
which more than one Higgs doublet couples to each type of fermion there will
in general be flavor-changing Yukawa interactions involving the
physical neutral Higgs fields~\cite{natfla}.  
There are stringent limits on such
couplings~\cite{fcnc}; for example, the $K_L - K_S$ mass difference
implies $ h/M_H < 10^{-6} GeV^{-1}$, where $h$ is the $\bar{d} s$
Yukawa coupling.
 
\vglue 0.6cm
\section{\elevenbf The Gauge Interactions}
\vglue 0.4cm
 
The major quantitative tests of the electroweak standard model involve
the gauge interactions of fermions and the properties of the gauge
bosons.  The charged current weak interactions of the Fermi theory and
its extension to the intermediate vector boson theory are incorporated
into the standard model, as is quantum electrodynamics.  The theory
successfully predicted the existence and properties of the weak
neutral current.  Here I will summarize the structure of the
interactions. Later chapters will discuss the phenomenology
and tests in more detail.
 
\vglue 0.6cm
\subsection{\elevenit The Charged Current}
\vglue 0.4cm
 
The interaction of the $W$ bosons to fermions is given by
\beq L = - \frac{g}{2\sqrt{2}} \left( J^\mu_W W^-_\mu + J^{\mu\dag
}_W W^+_\mu \right), \label{eqch49} \eeq
where the weak charge-raising current is
\begin{eqnarray}
J_W^{\mu \dag} &=& \sum^F_{m=1} \left[ \bar{\nu}_m^0 \gamma^\mu (1 -
\gamma^5) e^0_m + \bar{u}_m^0 \gamma^\mu (1-\gamma^5) d^0_m \right]
    \label{eqch50}    \\
&=& (\bar{\nu}_e \bar{\nu}_\mu \bar{\nu}_\tau ) \gamma^\mu (1 -
\gamma^5) \left( \begin{array}{c} e^- \\ \mu^- \\ \tau^- \end{array}
\right) + (\bar{u}\; \bar{c}\; \bar{t}) \gamma^\mu (1 - \gamma^5) V
\left( \begin{array}{c} d \\ s \\ b \end{array} \right)
.\nonumber  \end{eqnarray}
$J^{\mu \dag}_W$ has a $V-A$ form, \ie it violates parity and charge
conjugation maximally.  The mismatch between the unitary
transformations relating the weak and mass eigenstates for the up and
down-type quarks leads to the presence of the $F \x F$ unitary matrix $V =
A^{u \dag}_L A^d_L $  in the current.  This is the
Cabibbo-Kobayashi-Maskawa (CKM) matrix~\cite{cabkm}, which is ultimately due
to the mismatch between the weak and Yukawa interactions.  For $F = 2$
families $V$ takes the familiar form
\beq V = \left( \begin{array}{cc} \cos \theta_c & \sin\theta_c \\ -
\sin \theta_c & \cos \theta_c \end{array} \right), \label{eqch51} \eeq
where $\sin \theta_c \simeq 0.22$ is the Cabibbo angle.  This form
gives a good zero$^{\rm th}$-order approximation to the weak
interactions of the $u, d,s $ and $c$ quarks; their coupling to the
third family, though non-zero, is very small.
Including these couplings, the 3-family CKM matrix is
\beq V = \left( \begin{array}{ccc} V_{ud} & V_{us} & V_{ub} \\
V_{cd} & V_{cs} & V_{cb} \\ V_{td} & V_{td} & V_{td} \end{array}
\right), \label{eq330.10} \eeq
where the $V_{ij}$ may involve a CP-violating phase.

There is nothing to distinguish massless
neutrinos except their weak interactions, so one simply defines the
$\nu_e$ as the weak partner of the electron, and similarly for
$\nu_\mu$ and $\nu_\tau$.  If there were non-zero neutrino mass then
one would have to introduce a leptonic mixing matrix in the current, but its
effects would not be important in any process that is not actually sensitive
to the masses.
 
\begin{figure}
\small \def\baselinestretch{1} \normalsize
%\vspace{5cm}
\centering
\includegraphics*[scale=1.0]{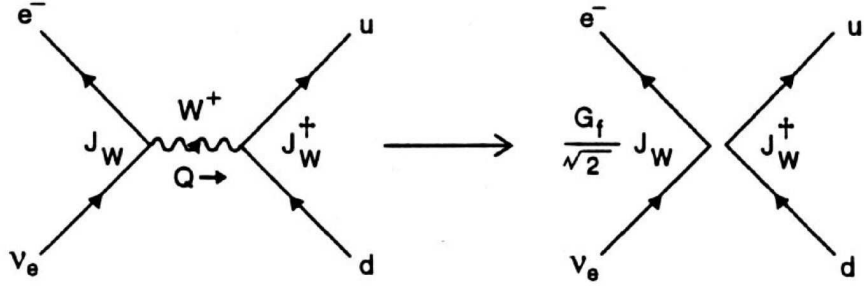}
\caption{A weak interaction mediated by the exchange of a $W$ and the
effective four-fermi interaction that it generates if the
four-momentum transfer $Q$ is sufficiently small.}
\label{figure12b}
\end{figure}
 
The interaction between fermions mediated by the exchange of a $W$ is
illustrated in Figure~\ref{figure12b}.  In the limit $|Q^2| \ll M_W^2$
the momentum term in the $W$ propagator can be neglected, leading to
an effective zero-range (four-fermi) interaction
\beq -L^{cc}_{\rm eff} = \frac{G_F}{\sqrt{2}} J^\mu_W J^{\dag}_{W\mu},
\label{eqch52} \eeq
where the Fermi constant is identified as
\beq \frac{G_F}{\sqrt{2}} \simeq \frac{g^2}{8M_W^2} =
\frac{1}{2\nu^2}.
\label{eqch53} \eeq
Thus, the Fermi theory is an approximation to the standard model valid
in the limit of small momentum transfer.
From the muon lifetime,  $G_F
= 1.16639(2) \x 10^{-5}$~GeV$^{-2}$, which implies that the weak
interaction scale defined by the VEV of the Higgs field is $\nu =
\sqrt{2} \langle 0| \varphi^0 |0 \rangle \simeq 246$~GeV.
 
The charged current weak interaction as described by (\ref{eqch52})
has been
successfully tested in a large variety of weak decays~\cite{wcc},
including $\beta$, $K$, hyperon, heavy quark, $\mu$, and $\tau$
decays.  In particular, high precision measurements of $\beta$, $\mu$, and
$\tau$ decays are a sensitive probe of extended gauge groups involving
right-handed currents and other types of new physics, as is described in
the chapters by Deutsch and Quin; Fetscher and
Gerber; and Herczeg. Tests of the unitarity
of the CKM matrix are important in searching for the presence of fourth
family or exotic fermions and for new interactions, as described by
Sirlin and by London. The standard theory has also
been successfully probed in neutrino scattering processes such as $ \nu_\mu
e \ra \mu^- \nu_e, \nu_\mu n \ra \mu^- p, \nu_\mu N \ra \mu^- X $. It
works so well that the neutrino-hadron interactions are used more as a probe
of the structure of the hadrons and QCD than as a test of the weak
interactions.
 
Weak charged current effects have also been observed in higher orders,
such as in the mass difference $M_{K_S} - M_{K_L}$, \CP\ violation in
the kaon system~\cite{cp}, and in $B \leftrightarrow \bar{B}$
mixing~\cite{bphys}.  For these higher order processes the full theory must
be used because large momenta occur within the loop integrals.
 
\vglue 0.6cm
\subsection{\elevenit QED}
\vglue 0.4cm
 
The standard model incorporates all of the (spectacular) successes of quantum
electrodynamics (QED)~\cite{QED}, which is based on the $U_{1Q}$ subgroup that
remains unbroken after spontaneous symmetry breaking.  The relevant
part of the \lag\ is
\beq L = - \frac{gg'}{\sqrt{g^2 + g^{\prime 2}}} J^\mu_{Q} ( \cos
\theta_W B_\mu + \sin\theta_W W^3_\mu), \label{eqch54} \eeq
where the linear combination of neutral gauge fields is just the
photon field $A_\mu$.  This reproduces the QED interaction provided
one identifies the combination of couplings
\beq e = g \sin \theta_W \label{eqch55} \eeq
as the electric charge of the positron, where $\tan  \theta_W \equiv
g'/g$. 
The electromagnetic current is given by
\begin{eqnarray}
J^\mu_{Q} &=&  \sum^F_{m=1} \left[ \frac{2}{3} \bar{u}^0_m
\gamma^\mu u^0_m - \frac{1}{3} \bar{d}^0_m \gamma^\mu d_m^0 -
\bar{e}^0_m \gamma^\mu e^0_m \right] \nonumber \\
           &=& \sum^F_{m=1} \left[ \frac{2}{3} \bar{u}_m
\gamma^\mu u_m - \frac{1}{3} \bar{d}_m \gamma^\mu d_m -
\bar{e}_m \gamma^\mu e_m \right].\label{eqch57} \end{eqnarray}
It takes the same form when written in terms of either weak
or mass eigenstates  because all fermions which mix with each
other have the same electric charge.
 Thus, the electromagnetic current is
automatically flavor-diagonal.

\vglue 0.6cm
\subsection{\elevenit The Neutral Current}
\vglue 0.4cm
 
The third class of gauge interactions is the weak neutral
current~\cite{wnc},
 which was predicted by the $SU_2 \x U_1$ model.
The relevant interaction is
\beq L = - \frac{\sqrt{g^2 + g^{\prime 2}}}{2} J^\mu_Z \left( - \sin
\theta_W B_\mu + \cos \theta_W W^3_\mu \right), \label{eqch59} \eeq
where the combination of neutral fields is the massive $Z$ boson field.  The
strength is conveniently rewritten as $ g/(2 \cos \theta_W)$, which
follows from $\cos \theta_W= g/\sqrt{g^2 + g^{\prime 2}}$.
 
The weak neutral current is given by
\begin{eqnarray} J^\mu_Z &=& \sum_m \left[\bar{u}^0_{mL} \gamma^\mu
u^0_{mL} - \bar{d}^0_{mL} \gamma^\mu d^0_{mL} + \bar{\nu}^0_{mL}
\gamma^\mu \nu^0_{mL} - \bar{e}^0_{mL} \gamma^\mu e^0_{mL} \right]
 -2 \sin^2 \theta_W J^\mu_{Q} \nonumber \\
&=&\sum_m \left[ \bar{u}_{mL} \gamma^\mu u_{mL} - \bar{d}_{mL}
\gamma^\mu d_{mL} + \bar{\nu}_{mL} \gamma^\mu \nu_{mL} - \bar{e}_{mL}
\gamma^\mu e_{mL} \right] 
 -2 \sin^2 \theta_W J^\mu_{Q}. \label{eqch60}
\end{eqnarray}
Like the
electromagnetic current $J^\mu_Z$ is flavor-diagonal in the standard
model; all fermions which have the same electric charge and chirality
and therefore can mix with each other have the same $SU_2 \x U_1$
assignments, so the form is not affected by the unitary
transformations that relate the mass and weak bases.  It was for this
reason that the GIM mechanism~\cite{gim} was introduced into the
model, along with its prediction of the charm quark.  Without it the
$d$ and $s$ quarks would not have had the same $SU_2 \x U_1$
assignments, and flavor-changing neutral currents would have
resulted. The absence of such effects is a major restriction on many
extensions of the standard model involving exotic fermions, as described in
the article by London.
The neutral current has two contributions. The first only involves the
left-chiral fields and is purely $V-A$.  The second is proportional to the
electromagnetic current with coefficient $\sin^2\theta_W$ and is purely
vector.  Parity is therefore violated in the neutral current interaction,
though not maximally.
 
\begin{figure}
\small \def\baselinestretch{1} \normalsize
%\vspace{5cm}
\centering
\includegraphics*[scale=1.0]{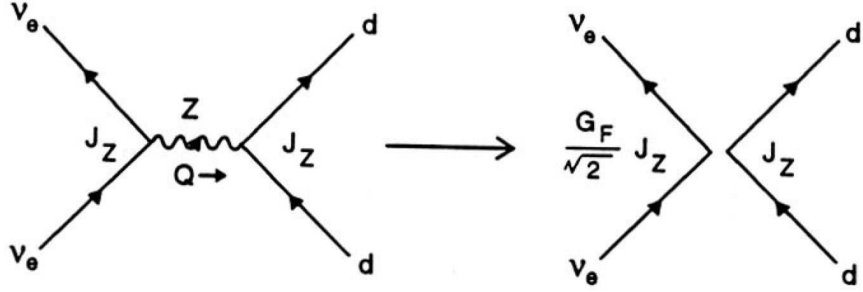}
\caption{Typical neutral current interaction mediated by the exchange
of the $Z$, which reduces to an effective four-fermi interaction in
the limit that the momentum transfer $Q$ can be neglected.}
\label{figure15b}
\end{figure}
 
In an interaction
between fermions in the limit that the momentum transfer is small
compared to $M_Z$ one can neglect the $Q^2$ term in the propagator,
and the interaction reduces to an effective four-fermi interaction
\beq - L^{NC}_{\rm eff}                           =
\frac{G_F}{\sqrt{2}} J^\mu_Z J_{Z\mu}.\label{eqch61} \eeq
The coefficient is the same as in the charged case because
\beq \frac{G_F}{\sqrt{2}} = \frac{g^2}{8 M_W^2} = \frac{g^2 +
g^{\prime 2}}{8M_Z^2}. \label{eqch62} \eeq
That is, the difference in $Z$ couplings compensates the difference in
masses in the propagator.  The weak neutral current was discovered at
CERN in 1973 by the Gargamelle~\cite{ggm} collaboration and by HPW at
Fermilab~\cite{hpw} shortly thereafter, and since that time it has
been extensively studied in many interactions, including $ \nu e \ra
\nu e, \; \nu N \ra \nu N, \; \nu N \ra \nu X$; $e^{\uparrow \downarrow} D
\ra e X$; atomic parity violation; $e^+ e^-$ and $Z$-pole reactions. These
have been the primary quantitative test of the unification part of the
standard electroweak model, and all aspects will be discussed extensively
in later chapters.
 
\vglue 0.6cm
\subsection{\elevenit Gauge Self-interactions}
\vglue 0.4cm
 
\begin{figure}
\small \def\baselinestretch{1} \normalsize
%\vspace{18cm}
\centering
\includegraphics*[scale=1.0]{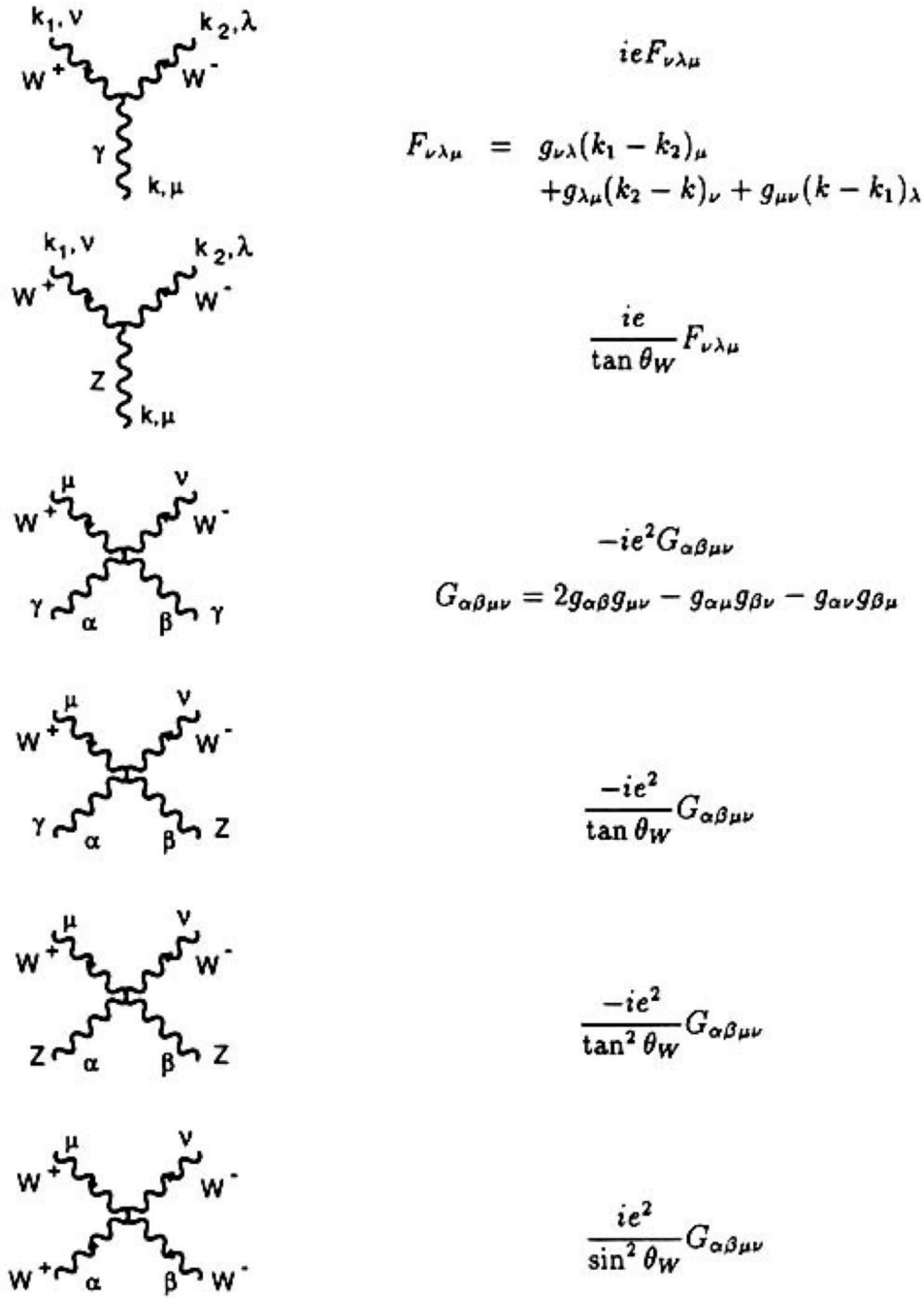}
\caption{The three and four point-self-interactions of gauge bosons in
the standard electroweak model.}
\label{figure16}
\end{figure}

\begin{figure}
\small \def\baselinestretch{1} \normalsize
%\vspace{4.0cm}
\centering
\includegraphics*[scale=1.0]{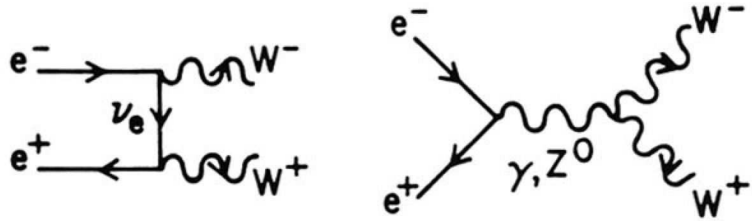}
\caption{Tree-level diagrams contributing to $e^+e^- \ra
W^+W^-$.}
\label{figure 3}
\end{figure}

The self-interactions of the gauge bosons in the standard
model are displayed in Figure~\ref{figure16}.  Their form is
predicted by the underlying gauge invariance, but they have not yet
been tested.  A sensitive test  will have to wait
for a study of  $e^+e^- \ra W^+W^-$ in the second phase, LEP~II, of the $e^+
e^-$ collider at CERN, as described  in the article 
by Treille~\cite{treille}, and future
possible  $e^+ e^-$ and hadron colliders at higher energy.
To lowest
order there are three diagrams, as shown in Figure~\ref{figure 3}.
Two of them involve the three-point interaction between a photon or
$Z$ boson and $W^+W^-$.  The cross section from any one of these
diagrams rises with center of mass energy, but gauge 
invariance relates these
three-point vertices to the couplings of the fermions in such a way that at
high energies there is a cancellation.  It is another manifestation of the
same cancellation which brings higher-order loop integrals under control,
leading to a renormalizable theory (otherwise, vector theories would have
severe divergences).   At LEP~II one will be able
to observe the cancellation; it would be even more
dramatic at possible future colliders at higher energies. 
Detailed studies of  $e^+e^- \ra
W^+W^-$ should be sensitive to deviations from the standard model,
especially those associated with such non-gauge physics as compositeness. In
practice, however, many of the types of new physics which could lead to
observable effects are already excluded by other observables at LEP~I and
elsewhere~\cite{effop}.
 
The processes $\stackrel{(-)}{q} q \ra V V'$ would also be sensitive
to gauge self-interactions.  Finally, one can study the
gauge-gauge three and four point vertices in the processes $e^+e^- \ra
e^+ e^- V V'$ and $\stackrel{(-)}{q} q \ra \stackrel{(-)}{q} q VV'$.  These
tests involve the same reactions that would be used to search for a very
heavy Higgs boson at a high energy hadron collider and will be important not
only for their own sake but as necessary background for the Higgs search.

\vglue 0.6cm
\section{\elevenbf Problems With the Standard Model}
\vglue 0.4cm
 
The \lag\ for the standard model after spontaneous symmetry breaking is
\begin{eqnarray} L &=& L_{\rm gauge} + L_{\rm Higgs}
   + \sum_i \bar{\psi}_i \left( i \not{\!\partial}- m_i - \frac{m_i
H}{\nu} \right) \psi_i \nonumber \\
&&- \frac{g}{2 \sqrt{2}} \left( J^\mu_W W^-_\mu + J^{\mu \dag}_W
W^+_\mu \right)
   - e J^\mu_{Q} A_\mu - \frac{g}{2\cos \theta_W} J^\mu_Z Z_\mu.
\label{eqch63} \end{eqnarray}

The standard electroweak model is a mathematically-consistent
renormalizable field theory which predicts or is consistent with all
experimental facts.  It successfully predicted the existence and form
of the weak neutral current, the existence and masses of the $W$ and
$Z$ bosons, and the charm quark, as necessitated by the GIM mechanism.
The charged current weak interactions, as described by the generalized
Fermi theory, were successfully incorporated, as was quantum
electrodynamics.  When combined with quantum chromodynamics for the
strong interactions and general relativity for classical gravity, the
standard model is almost certainly the approximately correct
description of nature down to at least $10^{-16}$cm, with the
possible exception of the Higgs sector.  However, the
theory has far too much arbitrariness to be the final story.  For
example, the minimal version of the model has 21 free parameters,
assuming massless neutrinos and not counting electric charge $(Y)$
assignments.  Most physicists believe that this is just too much for
the fundamental theory.  The complications of the standard model can
also be described in terms of a number of problems.

\vglue 0.3cm
\noindent 1. {\it Gauge Problem}
\vglue 0.2cm
 
\noindent
The standard model is a complicated direct product of three
sub-groups, $SU_3 \x SU_2 \x U_1$, with separate gauge couplings.
There is no explanation for why only the electroweak part is chiral
(parity-violating).  Similarly, the standard model incorporates but
does not explain another fundamental fact of nature: charge
quantization, {\it i.e.,} why all particles have charges which are
multiples of $e/3$.  This is important because it allows the
electrical neutrality of atoms $(|q_p| = |q_e|)$.  Possible
explanations include: grand unified theories~\cite{GUT}, the existence
of magnetic monopoles~\cite{monop}, and constraints from the
absence or cancellation\footnote{The absence of anomalies
is not sufficient to determine all of the $Y$ assignments without
additional assumptions, such as family universality.} 
of anomalies~\cite{anom}.
 
\vglue 0.3cm
\noindent
2. {\it Fermion Problem}
\vglue 0.2cm
 
\noindent
All matter under ordinary terrestrial conditions can be constructed
out of the fermions $ (\nu_e, e^-, u, d)$ of the first family.  Yet we
know from laboratory studies that there are $\geq 3$ families:
$(\nu_\mu, \mu^-, c, s)$ and $(\nu_\tau, \tau^-, t, b) $ are heavier
copies of the first family with no obvious role in nature.  (The $t$
and $\nu_\tau$ have not yet been directly observed, although there
are candidate $t$ events from CDF~\cite{CDF}. They are assumed to
exist because the weak interactions of the $b$ and $\tau$ have been
well measured and are in agreement with the assumptions that they
have $SU_2$-doublet partners~\cite{unique}.)  The standard model gives no
explanation for the existence of these heavier families and no
prediction for their numbers.  Furthermore, there is no explanation or
prediction of the fermion masses, which vary over at least five orders
of magnitude, or of the CKM mixings.  There are many possible
suggestions of new physics that might shed light on this, including
composite fermions; family symmetries; radiative hierarchies, in which
the fermion masses are generated at the loop-level~\cite{rad}, with
the lighter families requiring more loops; and the topology of extra
space-time dimensions, such as in superstring
models~\cite{strings}.  Despite all of these ideas there is no
compelling model and none of these yields detailed predictions.  The
problem is just too complicated.  Simple grand unified theories don't
help very much with this, except for the prediction of $m_b$ in terms
of $m_\tau$ in the simplest versions~\cite{mbmt}.
 
\vglue 0.3cm
\noindent
3. {\it Higgs/hierarchy Problem}
\vglue 0.2cm
 
\noindent
In the standard model one introduces an elementary Higgs field into
the theory to generate masses for the $W$, $Z$, and
fermions. For the model to be consistent the
Higgs mass should not be too different from the $W$ mass, \ie $M^2_H
= O (M_W^2) $. If $M_H$ were to be larger than $M_W$ by many orders of
magnitude there would be a hierarchy problem, and the Higgs
self-interactions would be excessively strong.  Combining theoretical
arguments with laboratory limits one obtains $M_H \lsim 1$~TeV.
(See (\ref{eqch39})).
 
\begin{figure}
\small \def\baselinestretch{1} \normalsize
%\vspace{6cm}
\centering
\includegraphics*[scale=1.0]{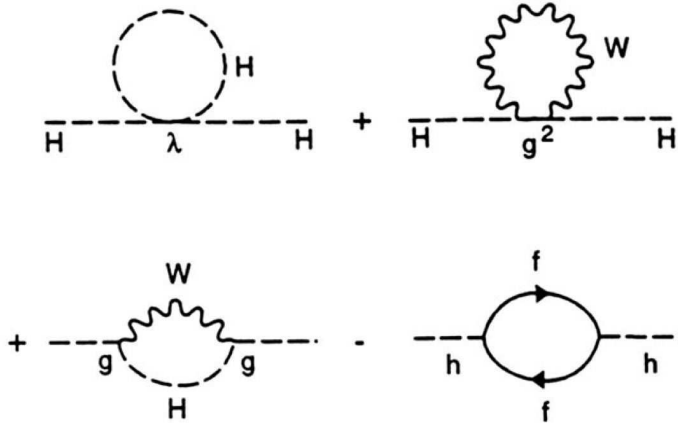}
\caption{Radiative corrections to the Higgs mass, including
self-interactions, interactions with gauge bosons, and interactions with
fermions.}
\label{figure20}
\end{figure}

However, there is a complication.  The tree-level (bare) Higgs mass
receives quadratically-divergent corrections from the loop diagrams in
Figure~\ref{figure20}.
One finds
\beq M_H^2 = (M_H^2)_{\rm bare} + O (\lambda, g^2, h^2) \Lambda^2,
\label{eqch64} \eeq
where $\Lambda$ is the next higher scale in the theory.  If there were
no higher scale one would simply interpret $\Lambda$ as an ultraviolet
cutoff and take the view that $M_H$ is a measured parameter and that
$(M_H)_{\rm bare}$ is not an observable.  However, the theory is
presumably embedded in some larger theory that cuts off the integral
at the finite scale of the new physics\footnote{There is no analogous
fine-tuning associated with logarithmic divergences, such as those
encountered in QED, because $\alpha \ln (\Lambda/m_e) < O(1)$ even
for $\Lambda = M_P$.}.  
For example, if the next
scale is gravity $\Lambda$ is the Planck scale $M_P = G_N^{-1/2} \sim
10^{19}$~GeV.  If there is a simple grand unified theory~\cite{GUT},
one would expect $\Lambda$ to be of order the unification scale $M_X
\sim 10^{14}$~GeV.  Hence, the natural scale for $M_H$ is
$O(\Lambda)$, which is much larger than the expected value.  There
must be a fine-tuned and apparently highly contrived
cancellation between the bare value and the correction, to more than
30 decimal places in the case of gravity.  If the cutoff is provided
by a grand unified theory there is a separate hierarchy problem at the
tree-level.  The tree-level couplings between the Higgs field and the
superheavy fields                         lead to the expectation
that $M_H$ is equal to the unification scale unless unnatural
fine-tunings are done.
 
One      solution  to this Higgs/Hierarchy problem is      the
possibility that the $W$ and $Z$ bosons are composite.  However, in
this case one would apparently be throwing out the successes of the
$SU_2 \x U_1$ gauge theory.
                                 Another approach is to eliminate
elementary Higgs fields in favor of a dynamical mechanism in which
they are replaced by bound states of fermions.
Technicolor and composite Higgs models are
in this category~\cite{technic}.  The third possibility is
supersymmetry~\cite{supersymmetry},
which    prevents large renormalizations by enforcing
                                   cancellations between the various
diagrams in Figure~\ref{figure20}.  However,
most grand unified versions do not explain why $(M_W/M_X)^2$ is so
small in the first  place.
 
\vglue 0.3cm
\noindent
4. {\it Strong \CP\ Problem}
\vglue 0.2cm
 
\noindent
Another fine-tuning problem is the strong \CP\ problem~\cite{stcp}.
One can add an additional term $\frac{\theta}{32 \pi^2} g^2_s F
\tilde{F}$ to the QCD \lag\ which breaks $P$, $T$ and \CP\ symmetry.
$\tilde{F}_{\mu \nu} = \epsilon_{\mu \nu \alpha \beta} F^{\alpha
\beta}/2$ is the dual field.
This term, if present, would induce an electric dipole moment $d_N$ for
the neutron.  The rather stringent limits on the dipole
moment~\cite{dipole} lead to the upper bound $\theta <
10^{-10}$.               The question is, therefore, why is $\theta$
so small?  It is not sufficient to just say that it is zero because
\CP\ violation in the weak interactions leads to a radiative
correction or renormalization of $\theta$ by $O (10^{-3})$.
Therefore, an apparently contrived fine-tuning is needed to cancel
this correction against the bare value.  Solutions include the
possibility that \CP\ violation is not induced directly by phases in
the Yukawa couplings, as is usually assumed in the standard model, but
is somehow violated spontaneously~\cite{stcp}.        $\theta$  then
would be a calculable parameter induced at loop level, and it is
possible to make $\theta$ sufficiently small.  However, such models
lead to difficult phenomenological and cosmological problems\footnote{Models
in which the \CP\ breaking occurs near the Planck scale may
be viable~\cite{barrnelson}.}.  Alternately, $\theta$ becomes unobservable
if there is a massless $u$ quark~\cite{masslessu}. However, most 
phenomenological estimates are not consistent with
$m_u = 0$~\cite{9L,leutwyler}.
Another
possibility is the Peccei-Quinn mechanism~\cite{PQ}, in which an extra
global $U_1$ symmetry is imposed on the theory in such a way that
$\theta$ becomes a dynamical variable which is zero at the minimum of
the potential.  Such models imply the existence of very light
pseudoscalar particles called axions.  Laboratory, astrophysical, and
cosmological constaints allow only the range $10^8 - 10^{12}$~GeV for
the scale at which the $U_1$ symmetry is broken.
 
\vglue 0.3cm
\noindent
5. {\it Graviton Problem}
\vglue 0.2cm
 
\noindent
Gravity is not fundamentally unified with the other interactions in
the standard model, although it is possible to graft on classical
general relativity by hand.  However, this is not a quantum theory,
         and there is no obvious way to generate one within the
standard model context.  In addition to the fact that gravity is not
unified and not quantized there is     another difficulty, namely the
cosmological constant.  The cosmological constant can be thought of as
energy of the vacuum.  The energy density                induced by
spontaneous symmetry breaking
is some 50 orders of magnitude larger than the observational upper
limit (see Eqns. (\ref{eqch41}) and (\ref{eqch42})).
This implies the necessity of severe fine-tuning between the
generated and bare pieces, which do not have any a priori reason to
be related.  Possible
solutions include Kaluza-Klein~\cite{KK} and supergravity
theories~\cite{supersymmetry}.  These unify gravity but do not solve the
problem of quantum gravity or yield renormalizable theories of quantum
gravity, nor do they provide any obvious solution to the cosmological
constant problem.  Superstring theories~\cite{strings} unify gravity
and may        yield finite theories of quantum gravity and all the
other interactions.  It is not clear whether or not they solve the
cosmological constant problem.


\begin{thebibliography}{999}
 
\small \def\baselinestretch{1} \normalsize

\bibitem{ss1} For reviews, see
E. S. Abers and B. W. Lee, \PR{9}{1}{73}; M. A. B. Beg and
 A. Sirlin, {\it ARNPS} {\bf 24}, 379 (1974),
 {\it Phys. Rep.} {\bf 88}, 1 (1982); P. Langacker, {\it Phys. Rep}. {\bf
 72}, 185 (1981) and in {\it TeV Physics}, ed. T. Huang \etal
 (Gordon and Breach, Philadelphia, 1991), p 53;
 {\it Testing the Standard Model}, ed. M. Cvetic and P.
 Langacker (World, Singapore, 1991);
P. Langacker and J. Erler, in {\it Reviews of Particle
  Properties}, \PR{D50}{1304}{94}.
\bibitem{ss1a}
H. Weyl, {\it Z. Phys.} {\bf 56}, 330 (1929);
C. N. Yang and R. Mills, \PR{96}{191}{54}.
\bibitem{QCD}  R. D. Field, {\it Perturbative QCD} (Addison-Wesley,
  Redwood City, 1989); Yu. L. Dokshitzer \etal {\it Basics of
  Perturbative QCD} (Ed. Frontieres, Gif-sur-Yvette, 1991);
  F. J. Yndurain, {\it The Theory of Quark and Gluon Interactions}
  (Springer-Verlag, Berlin, 1993).
\bibitem{ew} 
S. L. Glashow, \NP{22}{579}{61}; S. Weinberg,
\PRL{19}{1264}{67}; A. Salam in {\elevenit Elementary Particle Theory},
ed. N. Svartholm, (Almqvist and Wiksells, Stockholm, 1969) p 367.

\bibitem{ssb}  B. W. Lee, {\it Chiral Dynamics} (Gordon and Breach,
  NY, 1972); S. Coleman, {\it Aspects of Symmetry} (Cambridge Univ. Press,
    Cambridge, 1985).
\bibitem{J001} G. 't Hooft and M. Veltman, \NP{B50}{318}{72}, and
            references therein.
\bibitem{9h} S. Coleman and E. Weinberg, \PR{D7}{1888}{73}.
\bibitem{9e} P. W. Anderson, \PR{130}{439}{63};
P. W. Higgs, \PRL{12}{132}{64}, \con{13}{321}{64}, \PR{145}{1156}{66};
	F. Englert and R. Brout, \PRL{13}{321}{64};
       G. S. Guralnik, C. R. Hagen, and T. W. B. Kibble, \PRL{13}{585}{65},
        \PR{155}{1554}{67}.

\bibitem{9i} J. Goldstone, {\it Nuo. Cim.} {\bf 19}, 15 (1961);
Y. Nambu and G. Jona-Lasinio, \PR{122}{345}{61}, {\bf 124}, 246
(1961); Y. Nambu, \PRL{4}{380}{60}; J. Goldstone, A. Salam, and
S. Weinberg, \PR{127}{965}{62}.
\bibitem{lac29}  UA1: G. Arnison \etal \PL{B166}{484}{86};
   C. Albajar \etal \ZP{C44}{15}{89}; and references theirin.
\bibitem{lac30}  UA2: R. Ansari \etal \PL{B186}{440}{87};
J. Alitti \etal \PL{B276}{354}{92}; and references theirin.
\bibitem{9j} 
P. Janot, invited talk at {\it Neutrino 94}, Eilat, Israel,
 May 1994, Orsay LAL-94-59.
\bibitem{9f}
           J. F.~Gunion \etal   {\it The Higgs Hunters Guide}
           (Addison-Wesley, Redwood City, 1990); M.~Sher, {\it Phys. Rep.}
     {\bf 179}, 273 (1989);
M.S. Chanowitz, {\it ARNPS} {\bf 38}, 323 (1988) and in {\it TeV Physics}, p 1;
     H. Haber in {\it Testing the Standard Model}, p 340;
     {\it Perspectives on Higgs Physics}, ed. G. L. Kane (World, Singapore,
     1993).
\bibitem{coscon} For  reviews, see S. Weinberg, {\it Rev. Mod. Phys.} {\bf 61},
       1 (1989); M. J. Duncan, in {\it Testing the Standard Model}, p 743.
\bibitem{9L} C. A. Dominguez and E.~de~Rafael, {\it Ann. Phys.} {\bf 174},
           372 (1987); J.~Gasser and H.~Leutwyler, {\it Phys. Rep.} {\bf
           87}, 777 (1982); S.~Narison, \PL{B216}{191}{89}; J. F.
  Donoghue, {\it ARNPS} {\bf 39}, 1 (1989).
\bibitem{d0} D$\not{0}$: S. Abachi \etal \PRL{72}{2138}{94}.
\bibitem{CDF} CDF: F. Abe \etal \PRL{73}{225}{94}, \PR{D50}{2966}{94}.
\bibitem{natfla} S.L. Glashow and S. Weinberg, \PR{D15}{1958}{77}.
\bibitem{fcnc}  See, for example, L. Littenberg and G. Valencia,
   {\it ARNPS} \con{43}{729}{93}.

\bibitem{cabkm}
F. J. Gilman, K. Kleinknecht, and B. Renk, in {\it Reviews of Particle
  Properties}, \PR{D50}{1315}{94}.
\bibitem{wcc} For reviews, see G. Barbiellini and G. Santoni, {\it Riv.
Nuo. Cim.} {\bf 9} (2), 1 (1986); E. D. Commins and P. H. Buchsbaum, {\it
  Weak Interactions of Leptons and Quarks}, (Cambridge Univ. Press,
Cambridge, 1983).

\bibitem{cp}  {\it CP Violation}, ed. C. Jarlskog (World, Singapore, 1989).
\bibitem{bphys}  {\it B Decays}, ed. S. Stone (World, Singapore, 1992).


\bibitem{QED} {\it Quantum Electrodynamics}, ed. T. Kinoshita (World,
         Singapore, 1990).
\bibitem{wnc} 
{\elevenit Discovery
of Weak Neutral Currents: The Weak Interaction Before and After},
ed.  D. Cline and A. Mann,
AIP Conf. Proc. 300 (AIP, New York, 1994);
J. E. Kim \etal \RMP{53}{211}{81}; U. Amaldi \etal \PR{D36}{1385}{87}.
\bibitem{gim} S. L. Glashow, J. Iliopoulos, and L. Maiani, \PR{D2}{1285}{70}.
\bibitem{ggm} Gargamelle: F. J. Hasert \etal \PL{B46}{121, 138}{73}.
\bibitem{hpw}  HPW: A. Benvenuti \etal \PRL{32}{800, 1454, 1457}{74}.
\bibitem{treille}  See also 
{\it Physics at LEP}, ed. J. Ellis and R.  Peccei, Vol. 2, CERN 86-02.
\bibitem{effop}
A. De R\'ujula \etal \NP{B384}{3}{92};
C. P. Burgess and D. London, \PR{D48}{4337}{93};
C. P. Burgess \etal \PR{D49}{6115}{94}.
\bibitem{GUT} For reviews, see P. Langacker, {\it Phys. Rep.} {\bf C72}, 185
   (1981); {\it Ninth Workshop on Grand Unification}, ed. R. Barloutaud
(World, Singapore, 1988); G. G. Ross, {\it Grand Unified Theories},
 (Benjamin, 1985).
\bibitem{monop} See, for example, J. Preskill, {\it ARNPS}
{\bf 34}, 461 (1984).
\bibitem{anom} J. A. Minahan \etal \PR{D41}{715}{90}; C. Q. Geng and
   R. E. Marshak, \PR{D41}{717}{90}; C. Q. Geng, \PR{D41}{1292}{90}; K.S.
  Babu and R. Mohapatra, \PR{D41}{271}{90}; S. Rudaz, \PR{D41}{2619}{90}.
\bibitem{unique} 
P. Langacker, {\elevenit Comm. Nucl. Part. Sci.} {\elevenbf 19}, 1
(1989); D. Schaile and P. M. Zerwas, \PR{D45}{3262}{92}.
\bibitem{rad} See, for example, K. S. Babu and R. Mohapatra,
   \PRL{66}{556}{91}; B. S. Balakrishna \etal \PL{205B}{345}{88}.
\bibitem{strings} M. B. Green, J. H. Schwarz, and E. Witten,
{\it Superstring Theory}, (Cambridge Univ. Press, Cambridge, 1987).
\bibitem{mbmt}
M. S. Chanowitz, J. Ellis and M. K. Gaillard,
{\it Nucl. Phys.} {\bf B128}, 506 (1977);
A. J. Buras, J. Ellis, M. K. Gaillard and
D. V. Nanopoulos, {\it ibid.} {\bf 135}, 66 (1978).
\bibitem{technic} For a review, see T. Appelquist, in
{\it Mexican School of Particles and Fields}, 1990, (QCD161:M45:1990), p 1.
\bibitem{supersymmetry}
See, for example, H. P. Nilles in {\it Testing the Standard Model}, p 633 and
Phys. Rep.
{\bf C110}, 1 (1984); H. E. Haber and G. Kane, {\it Phys. Rep.}
{\bf C117}, 75 (1985).
\bibitem{stcp}  R. Peccei, in {\it CP Violation}, p 503.
\bibitem{dipole} {\it Reviews of Particle Properties},
L. Montanet \etal \PR{D50}{1180}{94}.
\bibitem{barrnelson}  A. Nelson, \PL{136B}{387}{83}, \con{143B}{165}{84};
    S. Barr, \PR{D30}{1805}{84}, \con{D34}{1567}{86}.
\bibitem{masslessu}   D. B. Kaplan and A. V. Manohar, \PRL{56}{2004}{86}.
\bibitem{leutwyler}  H. Leutwyler, \NP{B337}{108}{90}, Bern BUTP-94/8.
\bibitem{PQ} R. D. Peccei and H. R. Quinn, \PRL{38}{1440}{77},
 \PR{D16}{1791}{77}; S. Weinberg, \PRL{40}{223}{78};
  F. Wilczek, \PRL{40}{271}{78}.
\bibitem{KK}   {\it Modern Kaluza-Klein theories}, ed. T. Appelquist,
 A. Chodos, and P. G. O. Freund, (Addison-Wesley, Menlo Park, 1987).
   \end{thebibliography}
\end{document}